\documentclass[aps,twocolumn,superscriptaddress]{revtex4}
\usepackage{amsmath,amssymb}
\usepackage{graphics,graphicx}
\usepackage{dcolumn,bm}
\usepackage{psfrag}
\usepackage{color}
\usepackage{multirow}
\newcommand{\cu}
{\affiliation{Department of Physics, University of Calcutta,
92 Acharya Prafulla Chandra Road, Kolkata 700009, India.}}

\newcommand{\victo}
{\affiliation{Department of Physics, Victoria Institution (College),
78B Acharya Prafulla Chandra Road, Kolkata 700009, India.}}

\begin{document}

\title{Non-equilibrium dynamics in a  three state opinion formation model with stochastic extreme switches}
 
\author{Kathakali Biswas}
\victo
\cu
\author{Parongama Sen}
\cu

\begin{abstract}
We investigate the  non-equilibrium dynamics of a  three state  kinetic exchange model of opinion formation, where switches between extreme states are possible, 
depending on the value of a  parameter 
$q$. 
 The mean field dynamical equations are derived and analysed for any $q$. The fate of the system under the evolutionary rules used in 
S. Biswas et al,  
Physica A {\bf 391}, 3257 (2012)
shows that it is dependent on the value of $q$ and the initial state in general. 
For  $q=1$, which allows the extreme switches maximally,  
a quasi-conservation in the dynamics is obtained which renders it equivalent to the  voter model. For general  $q$ values, a ``frozen''  disordered fixed point is  obtained which acts as  an attractor for all initially disordered states. 
For other initial states,   the order parameter grows with time $t$ as  $\exp[\alpha(q) t]$ where 
$\alpha = \frac{1-q}{3-q}$ for $q\neq 1$ and follows  a power law behaviour for $q=1$.
 Numerical simulations using a fully connected 
agent based model provide additional  results like  the system size dependence of the  exit probability and consensus times  that further accentuate  the different behaviour of the model for $q=1$ and $q\neq 1$.  
The results are 
compared with the non-equilibrium phenomena in other 
well known  dynamical systems.

\end{abstract}

\maketitle

\section{Introduction}
One of the main motivations  in studying non-equilibrium phenomena is to check what kind of steady states  can be reached using different initial conditions. In the well known studies of Ising-Glauber  model at zero temperature, on lattices or networks, several studies have been made to show that the 
steady states may not be the equilibrium steady states \cite{spirin1,spirin2,barros,svenson,hagg,boyer,castellano,biswas,baek,khaleque,krap-comp,castell2003,sood-red2005,such-2005-1}. Exit probability,  a quantity related to the type of 
final state  reached from an initially biased state, has also been studied extensively in recent times in spin and opinion formation models \cite{claudio,pkm,sb_ps,pr_sb_ps,sb_ps_pr,prado,pr_ps15,timp,pm_ps16,pr_ps17,pm_ps17,sm_sb_ps}. In  systems with more than two states,
several other interesting features like two stage ordering process has been noted \cite{sm_sb_ps}. 
In addition, how a system evolves to a stable state starting from an unstable fixed point is also a matter of interest \cite{RRPS}.

 Opinion dynamics models relevant to social phenomena have received extensive attention recently \cite{stauffer,soc_rmp,sen_chak,galam_book}. 
{ These models typically show a rich non-equilibrium behaviour. Usually,  the opinion of an agent is updated following the interaction with other individuals; sometimes the influence of media is also incorporated. In the numerous models studied so far,
the interaction and the choice of the interacting agent(s) play  crucial roles.  
The simplest models involve binary opinions typically represented by 0,1 or $\pm 1$. The Voter 
model \cite{voter,ligg}, in which an  agent just copies the opinion of another randomly
picked up agent, is one  of the simplest and earliest  opinion dynamics  models.
 Later, 
models involving   
more complexities have been constructed \cite{soc_rmp,sen_chak}.   The  binary models obviously cannot capture all the 
intricacies  of the real world.
Hence,   models with 
three or more opinion states as well as continuous values of opinions have been considered in the recent past. 
The voter model can be generalised with more number of states easily \cite{starnini} while other multistate models which involve the effect of more neighbours have also been considered \cite{Szolo,viela}. In comparison to the simple binary state models, here the opinions 
are not merely flipped but can change in more than one possible way. 
We focus our attention  on  the so called  kinetic exchange models  where pairwise interactions are considered at each step \cite{toscani}.     However, these models generally have some restrictions.}
In particular, in the kinetic 
exchange models most recently studied with three discrete opinion states quantified  by -1,0,1 (assumed to represent e.g., left, central and right ideologies), 
a  transition from 1 to -1 or vice versa (i.e., 
an extreme switch of opinion)  is  not allowed to the best of our knowledge 
\cite{BCS,meanfield,nuno1,nuno2,sudip}.
{ Also, in many other similar three-state models such a restriction is imposed
\cite{vazquez,vazquez2,mobilia,lima,migu,cast,luca}.}
However, human behaviour being complex and unpredictable such switches cannot be completely ruled out. In fact, there are  real world examples where even political cadres or leaders  shift their  allegiance to  parties with totally opposite principles \cite{tele,express}. The reasons may be associated with immediate gains and selfish interests, lack of strong ideological beliefs etc.  
{ 
We have  considered a model for opinion dynamics where  extreme switches are allowed to happen and see how the dynamics are affected by this.
It may be added here  that for the multistate voter model or Potts type models, 
such extreme switches can  
take place, however,  in the  relevant studies, the effect of such switches has not been the 
issue of interest  specifically \cite{starnini,Szolo,viela}.} 

In this article, we have  considered
 a kinetic exchange model of opinion dynamics with three states, with the possibility of switching between extreme opinions. 
In the mean field approach, the equations for the 
time derivatives are set up for the three population densities  of different opinions and solved numerically.  
 We have introduced a parameter $q$ which  governs  the  probability with which switches  between extreme opinions can occur and studied its effect on the time 
evolution. $q$ varies between zero and unity, the zero case is already considered 
where no such switch is allowed \cite{BCS}. Parallely, numerical simulations have been conducted using a fully connected agent based model.   
The model and quantities of interest are discussed in the next section followed by the results presented in section \ref{results} and
finally in the  concluding section, the results are discussed and compared to existing results 
in similar models.

\section{Mean Field Kinetic Exchange Model}

 We have considered a  kinetic exchange model (KEM) for opinion formation  which incorporates  three opinion values   are quantified by $0, \pm 1$. The possible correspondence with left, central and right ideologies has already been mentioned. The three opinion values may even 
mimic a     2-party voting system, where the the $\pm 1$ opinions represent   support for the two parties while people with  
zero opinion (the neutral population) are those who refrain from voting for either of them. 
The opinion of an  individual is updated by taking into account her  present opinion and an interaction with a randomly chosen 
 individual in the fully connected model. 
The opinion   of the $i$th  individual  is  denoted by 
$o_i(t)$.
The time evolution of $o_i$, after an interaction 
   with the $k$th individual, chosen randomly, is given by 
 
\begin{equation}
o_i(t+1)=o_i(t) + \mu  o_k(t),
\end{equation}
where $\mu$ can be interpreted as an interaction parameter. 
The opinions are bounded in the sense $|o_i| \leq 1$ at all times and therefore 
$o_i $ is taken as 1 (-1) if it is more  (less) then 1 (-1). There is no self-interaction so $i \neq k$ in general.  
This evolutionary rule was introduced in \cite{BCS}. 
Here time is assumed to be discrete but one can easily use a continuous time model as will be done in this paper. 

\begin{center}
\begin{figure}[h]
\centering
\includegraphics[width=9cm, keepaspectratio]{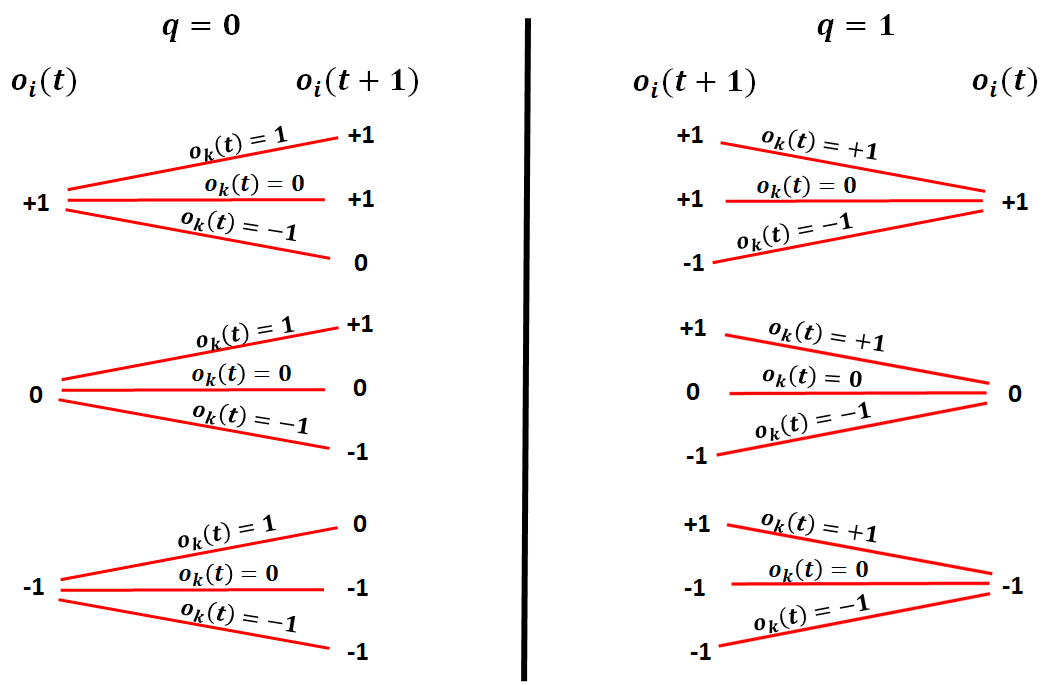}
\caption{The  updated opinions of the $i$th individual following an interaction with another 
individual 
(denoted by $k$) 
for all possible opinion values at time $t$ are shown for $q = 0$ (left panel), which implies $\mu = 1$    and $q=1$ (right panel) for which  $\mu = 2$. }
\label{fig1:inter} 
\end{figure}
\end{center}

 In several previous works \cite{BCS,sm_sb_ps,meanfield,nuno1,nuno2,sudip}, $\mu$, the interaction parameter,  has been chosen randomly, allowing also negative values albeit being bounded; $|\mu| \leq 1$.  
Such a bound  allows a transition between opinion values with a difference of maximum $\pm 1$ only. 
In the present  work, the interaction parameter $\mu$ is allowed to take two discrete values. 
The values are $\mu = 1$ and $\mu = 2$  which occur  with  probabilities  $1-q$ and  $q$ respectively.  Hence, for example, 
if an agent with opinion $+1$ interacts with another with opinion $-1$ and $\mu =2$, her  opinion can change to -1, the other extreme value. The possibilities of all the interactions and resulting opinions are shown in Fig.\ref{fig1:inter} for the extreme values $q=0$ and $q=1$.  
Note that in the present work, only positive values of $\mu$ are allowed.

The densities of the three populations with opinion $0, \pm 1$ are 
denoted by  $f_0, f_{\pm 1}$ with $f_0 + f_{+1} + f_{-1} = 1$. The ensemble averaged order parameter
obtained from the time dependent equations for the densities
is given as $\langle O(t) \rangle = f_{+1} - f_{-1}$ with $-1 \leq \langle O(t) \rangle \leq  1$. 

Usually,  to study the 
 opinion dynamics models, one  starts with 
a random disordered configuration  such that the average opinion is $~0$. Given that there are three states, one can choose this state with different combinations of $f_i'$s,  keeping 
$f_{+1} = f_{-1}$. 
A conventional choice is $f_0 = f_{\pm 1} = 1/3$.

One can also study  the effect of an  initial bias in the distribution of opinions in the starting configuration of the system. The homogeneous 
configuration being one with all the densities equal to 1/3,
one can consider a deviation from this such that the net opinion is nonzero
by choosing  $f_0= 1/3$, $f_{+1}= 1/3 + \Delta/2$  and  $f_{-1}=1/3 - \Delta/2$. 
Here $-2/3 \leq \Delta \leq 2/3$.
Apart from this case, one can take other initial configurations  which have a net nonzero opinion. We have discussed such cases as well to show the initial configuration dependence. 

We present in this paper the rate equations derived analytically using mean field 
theory for the three densities, and study their behaviour  as  functions of time.
The fixed point analysis of the equations present some interesting and non-intuitive results. 
We also obtain the  exit probability. 
Here the exit probability $E$ is considered as a function of $\Delta$, i.e., 
$E(\Delta) $ is the probability that the final configuration has $f_{+1} =1$ starting from $f_{+1} = 1/3 + \Delta /2$ and $f_{-1} = 1/3 - \Delta /2$. 
The saturation value of  $\langle O\rangle$ is related to $E(\Delta)$ by 
\begin{equation}
\langle O \rangle _{sat}= 2E(\Delta) - 1
\label{satexit}
\end{equation}
from which the exit probability can be estimated. 
 
We have also conducted numerical simulations by considering an agent based model where each agent can interact with any other agent. 
Here, the order parameter for a given configuration is 
defined as $\bar O(t) = \frac{|\sum o_i(t)|}{N}$ where $N$ is the system size with  $\langle \bar O\rangle$ denoting the configuration average. To calculate the exit probability 
$E(\Delta)$, we directly estimate the fraction of   configurations which reach the consensus state with all opinions equal to 1. 

To solve the coupled differential equations, Euler method has been used and 
in the Monte Carlo method, system sizes ranging from 100 to $2^{16}$ have been 
simulated  with number of configurations ranging between $10^4$ to $10^5$. 

From the simulations, it is also possible to estimate the average consensus times for different system sizes. All the results are presented in the next section.

\section{Results }
\label{results}

We present in this section   the mean field analytical solution in detail and also the results obtained from numerical simulations.


\subsection{Mean field rate equations}

To set up the rate equations for the $f_i$'s, we need to treat the time variable as continuous. 
   Assume that the opinion changes from $i$ to $j$ ($i,j = 0,\pm1$)  
in time $\Delta t$  with the transition rate given by     $w_{i \rightarrow j}$.
Then we have the following set of $w_{ij}$'s:  
\begin{eqnarray*}
w_{+1 \rightarrow +1} &  =& f^2_{+1} + f_0f_{+1}\\
w_{0 \rightarrow +1} & = & f_0f_{+1}  \\
w_{-1 \rightarrow +1} & =&  qf_{-1}f_{+1} \\
w_{+1 \rightarrow 0}  & = & (1-q)f_{+1}f_{-1}\\
w_{0 \rightarrow 0} &  = & f^2_0\\
w_{-1 \rightarrow 0} &  = & (1-q)f_{-1}f_{+1}\\
w_{+1 \rightarrow -1} &  = & qf_{+1}f_{-1}\\
w_{0 \rightarrow -1} &  =&  f_0f_{-1}\\
w_{-1 \rightarrow -1} &  =&  f^2_{-1} + f_0f_{-1}
\end{eqnarray*}


Hence, in general, we have $f_i(t+\Delta t) = f_i(t) + \sum_j w_{j\to i} \Delta t - \sum_j w_{i \to j}\Delta t$ 
such that taking $\Delta t \to 0$, we get   
\begin{equation}
\frac{df_{+1}}{dt}  
=  f_0f_{+1} - (1-q)f_{+1}f_{-1},
\label{f+equation}
\end{equation}
and
\begin{equation}
\frac{df_{-1}}{dt} 
=   f_0f_{-1} -(1-q)f_{-1}f_{+1}.
\label{f-equation}
\end{equation}

%
%
The time evolution of 
the ensemble averaged order parameter $\langle O(t)\rangle$  satisfies

\begin{equation}
\frac{d\langle O(t)\rangle}{dt} = f_0\langle O(t)\rangle.
\label{Oequation}
\end{equation}

%

\begin{figure}
\includegraphics[width=8cm]{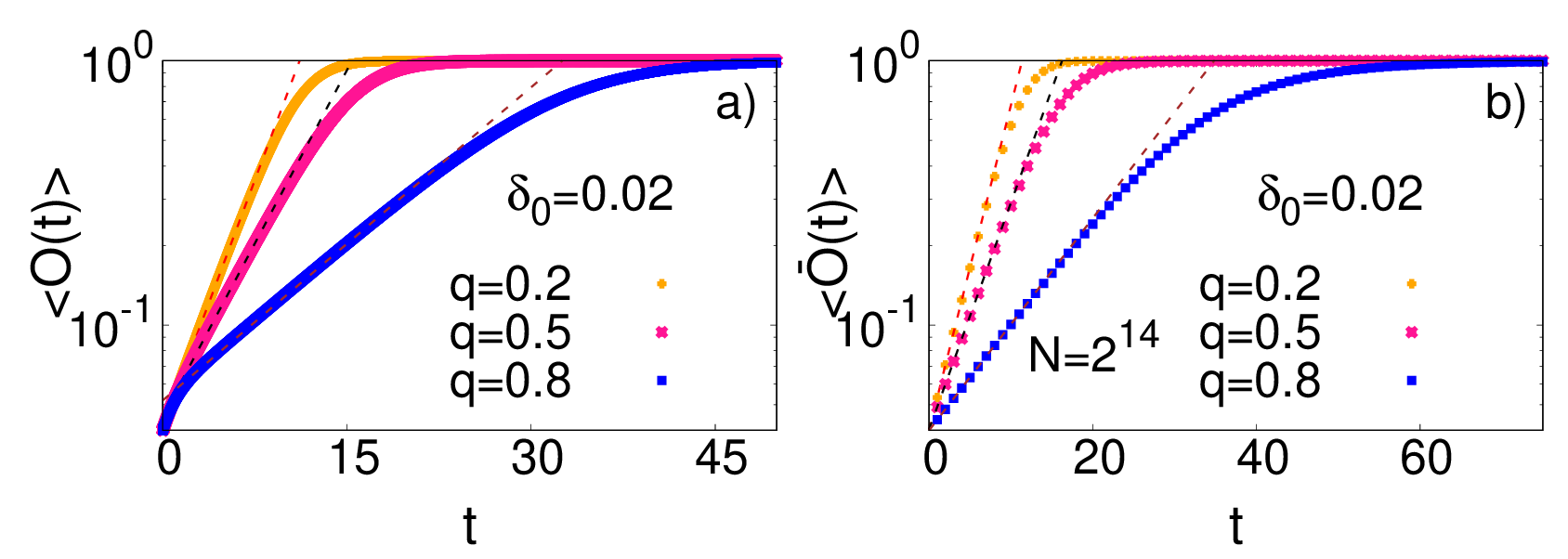}
\caption{The order parameter versus time variation
  near the frozen fixed point shows an exponential growth given by  $\exp[\alpha (q)t]$ for any $q \neq 1$  
as is evident from the (a) analytical as well as (b) simulation results. 
Also shown are the  best fitted curves of the form given in  eq. \ref{delta}.}
\label{orderpara1}
\end{figure}

\subsection{Fixed points and steady states}

There will be some trivial fixed points corresponding to the initial conditions 
that have  any of the three densities  equal to 1. Here, obviously, there will be no evolution of the system at all. We consider more general cases in the following. 

Equation \ref{Oequation} shows that a steady state   for $\langle O\rangle$ is obtained when   $\langle O\rangle  = 0$ and/or $f_0 = 0$.
Consider first the case   when we have a  disordered steady state, i.e., $\langle O (t \to \infty) \rangle   = 0$. 
If the initial state is disordered, eq. \ref{Oequation} indicates that $O$ will remain zero, i.e.,
will not evolve although the individual densities may change in time. 
We  show in the following that  for    all values of $q$, there  exists a 
non-trivial disordered fixed point at which there is  no evolution of not only  the order parameter but also of the individual densities.
This special fixed point may be termed the frozen fixed point (FFP) as the system 
does not undergo any change at all right from the beginning, although none of the 
densities have value unity.

At the FFP, 
  $f_{+1} = f_{-1} = x > 0 $ is a constant in time.  
Using this in eq. \ref{f+equation} or \ref{f-equation}, one gets
\begin{equation}
\dfrac{dx}{dt} = x - (3-q)x^2  = 0.
\end{equation}

Ignoring the  solution $x =0$, 
we get   $x=\dfrac{1}{3-q}$,  i.e., the fixed point is given by   
\begin{equation}
    f_{+1}=f_{-1}=\dfrac{1}{3-q}; f_0=\dfrac{1-q}{3-q}.
\label{FFP}
\end{equation}
%

The stability of the FFP can be checked by introducing small deviations about these
values. These deviations can be introduced in different ways. 
We first consider a deviation such that the initial state has a nonzero order. 
Taking  
$f_{+1} (t) =  x^* + \delta $ and 
$f_{-1} (t) =  x^* - \delta $  
where $x^* = \frac{1}{3-q}$ is the FFP value, we get from 
eq. \ref{f+equation}
\begin{gather*}
\dfrac{d(x^*+\delta)}{dt}=(x^*+\delta) - (2-q)(x^*+\delta)(x^*-\delta) - (x^*+\delta)^2.
\end{gather*}

Linearising the above, one finally gets
\begin{equation}
\delta (t)=\delta _0 \exp[\alpha (q)  t],
\label{delta}
\end{equation} 
where 
 $\delta_0$ is the initial value of $\delta$ and 
\begin{equation}
\alpha (q)= \dfrac{1-q}{3-q}.
\label{alpha}
\end{equation}
The  order parameter which is equal  to $2\delta(t)$ should show the same behaviour 
and indeed 
both the analytical solution and simulations show the expected initial exponential growth with the value of the 
exponent very close to $\alpha$ given by eq. \ref{alpha} (see Fig. \ref{orderpara1}).   The
simulation results  have some finite size effects which is not unexpected (Fig. \ref{orderpara1}b). 
The fact that $\alpha (q) > 0$ (for $q \neq 1$) implies the FFP is an unstable one for all values of $q$ (except $q=1$) when the deviation  favours a finite order.

On the other hand, if we start from any disordered state, it can be shown that the system   
  will flow towards the FFP. Here, with $f_{+} = f_{-1}$ initially, they will remain the same 
in time as indicated by the rate equations. 
 Hence  the   state can be characterised by  $f_\pm = x^ * + \rho$ and $f_0 = 1-2(x^* +\rho)$. 
In this case, we obtain 
\begin{equation}
\rho (t) = \rho_0  \exp[- t],
\end{equation} 
i.e., the state flows to the FFP with a rate independent of $q$. Here $\rho_0$ is the initial value of $\rho$.
Hence the FFP acts as an attractor for all initially disordered state. 
We have checked that the above form is indeed obeyed for any value of $q$ (not shown).

\subsection{Time evolution of the densities and the order parameter}

Having obtained the fixed point and the behaviour of the system close to it for any value of $q$, we proceed to study 
the time evolution of the relevant variables 
in more detail in this subsection. 
We will first discuss this for   $q=1$, which is obviously a special point in the parameter space.  
For other values  of $q$ also, we present  the  results which show consistency with the theoretical analysis. 
The data for the time evolution of the three densities and the order parameters  have been  obtained
 by numerically solving the analytical equations  and also using Monte Carlo  simulations. 
 Only two of these four quantities are 
independent, however,  it is more informative to present the results for all of them.


\begin{figure}[h]
\centering
\includegraphics[width=8cm,angle=0]{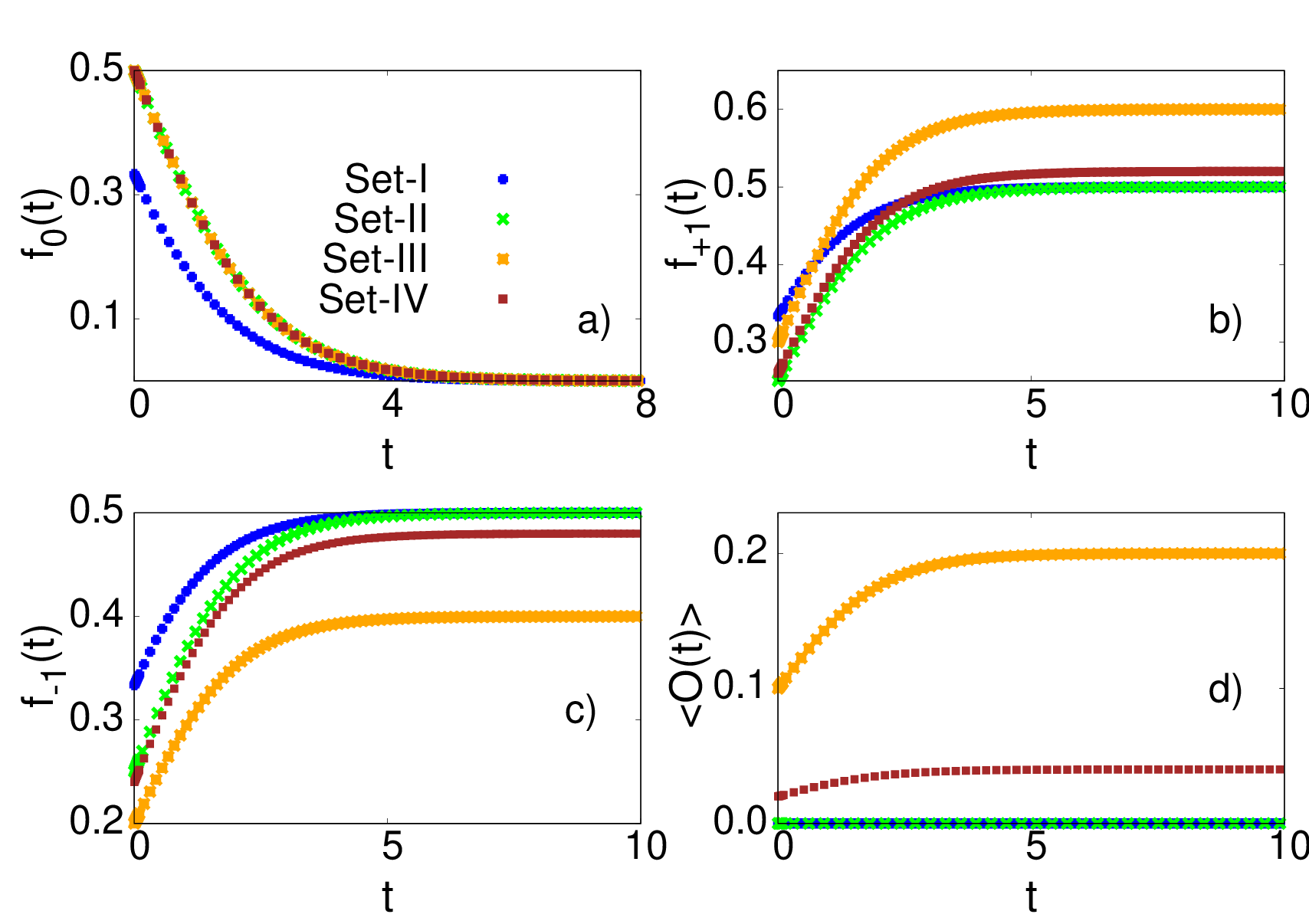}
\caption{Results for $q=1$ 
obtained from the analytical solution using the initial configurations given  in Table \ref{table:1}.
The three densities  and the ensemble averaged order parameter are  
shown  as functions of time in  (a), (b), (c) and (d) respectively. The time evolution of sets I and II merge within a few steps as expected.} 
\label{fig:q=1,0-case-1-set-pic} 
\end{figure}

\begin{figure}[h]
\centering
\includegraphics[width=8cm,angle=0]{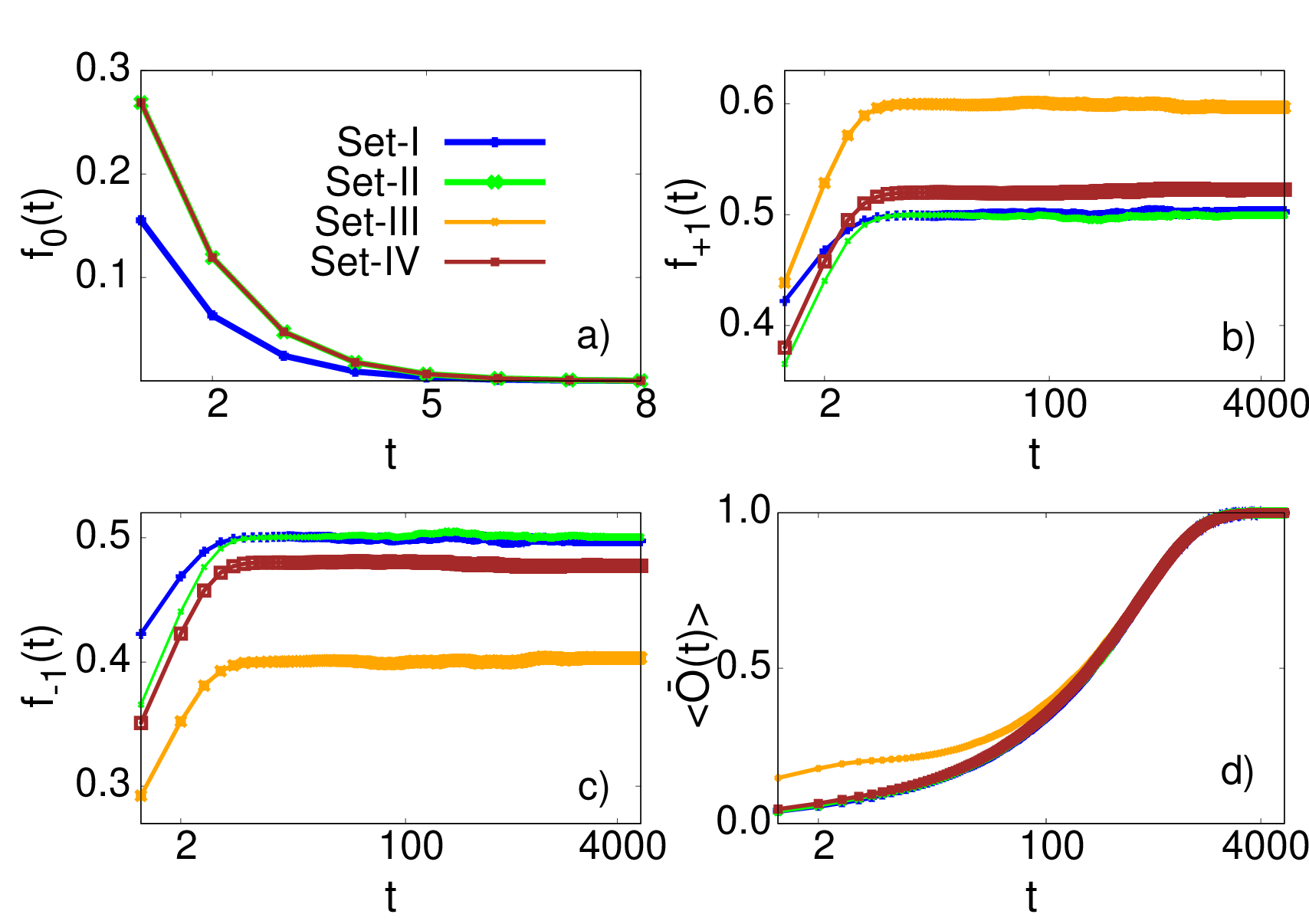}
\caption{Results for $q=1$ 
obtained from  Monte Carlo simulation (of system size $N=2^{10}$ )  using the initial configurations given  in Table \ref{table:1}.
The three densities  and the ensemble averaged order parameter are  
shown  as functions of time in  (a), (b), (c) and (d) respectively.} 
\label{fig:q=1,0-case-1-set-simulation-pic} 
\end{figure}

 We have used four different sets of initial conditions stated in Table \ref{table:1}. Of these,  sets I and II are both disordered with   set I corresponding  to the homogeneous case. Set III represents an arbitrary initial condition that favours order. Set IV is also ordered initially and can be regarded as a small deviation from set II. 
For all these cases,  for $q=1$, 
$f_0$ falls  rapidly 
within a few steps  
as shown in Fig.~\ref{fig:q=1,0-case-1-set-pic}a and Fig.~\ref{fig:q=1,0-case-1-set-simulation-pic}a
 obtained using both the methods.   
This behaviour of $f_0$ may be easily understood from the transition possibilities, as we note (see Fig. \ref{fig1:inter}) that for opinion zero, there is no flux to this state 
from  opinions values $\pm 1$ while there is an outgoing flux  when the  zero  opinion 
changes to other values.
This leads to the behaviour of $\langle O(t)\rangle$ in eq. \ref{Oequation} as 
$\frac{d\langle O\rangle }{dt}   \approx 0$,
i.e., a quasi-conservative system is obtained. 
Note that $f_0 \to 0$ implies $f_{\pm 1}$ are independent of time for $q=1$, but not necessarily equal to 1 or 0. 
Hence, the consensus state (i.e., either $f_{+1}$ or $f_{-1}$ equal to 1) is not reached  
in general such that the value of the ensemble averaged 
opinion is less than 1. We exclude here the trivial cases 
where $f_{+1}=1 $ or $f_{-1}=1$ initially.  
The results using the mean field equations   are shown in  
Fig. \ref{fig:q=1,0-case-1-set-pic} for different initial states given in Table \ref{table:1}.  
It is seen that as expected,   for sets I and  II, which are disordered initially,  
the system evolves to the  FFP $0, 1/2, 1/2$. For the other sets we see that  
the system reaches a steady state (which is not a consensus state) within a few steps
with the final value of the order parameter close to the initial one and $f_0=0$. 
For sets III and IV, we have used initial states 
with a bias to the +1 opinion and  $O(t)$ is therefore positive in all the cases. 


The corresponding simulation results are shown  in  
Fig. \ref{fig:q=1,0-case-1-set-simulation-pic}. Here the consensus states are reached for all the
sets of initial states including the disordered ones (sets I and  II). 
 This is because, in simulations, since  we have a  finite system size, a   random fluctuation can drive the system to a consensus state in an individual configuration.
Therefore  the data which are shown for  the ensemble average of the {\it absolute value} of the order parameter shows $\langle \bar O\rangle  \to 1$ at large times for all initial states. 
This is analogous  to the kinetics in Ising Glauber model at zero temperature in one dimension, where  we have a conservation such that the ensemble averaged 
magnetisation is zero. 
In simulations, however, an individual configuration  indeed reaches the all spin up/down state 
 so that the absolute value of magnetisation reaches  unity even after configuration 
averaging.


\begin{table}[h]
\begin{center}
\begin{tabular}{ |c|c|c|c| }
 \hline
 Initial configuration & $f_0$ & $f_{+1}$ & $f_{-1}$\\
 \hline
 Set-I & 1/3 & 1/3 & 1/3 \\
 Set-II & 1/2 & 1/4 & 1/4 \\
 Set-III & 5/10 & 3/10 & 2/10 \\
 Set-IV& 1/2 & 1/4 + 0.01 & 1/4 - 0.01\\
 \hline
\end{tabular}
\end{center}
\caption{Fraction of neutral opinion, positive opinion and negative opinion considered for the initial configuration.  }
\label{table:1}
\end{table}


\begin{figure}[h]
\centering
\includegraphics[width=8cm,angle=0]{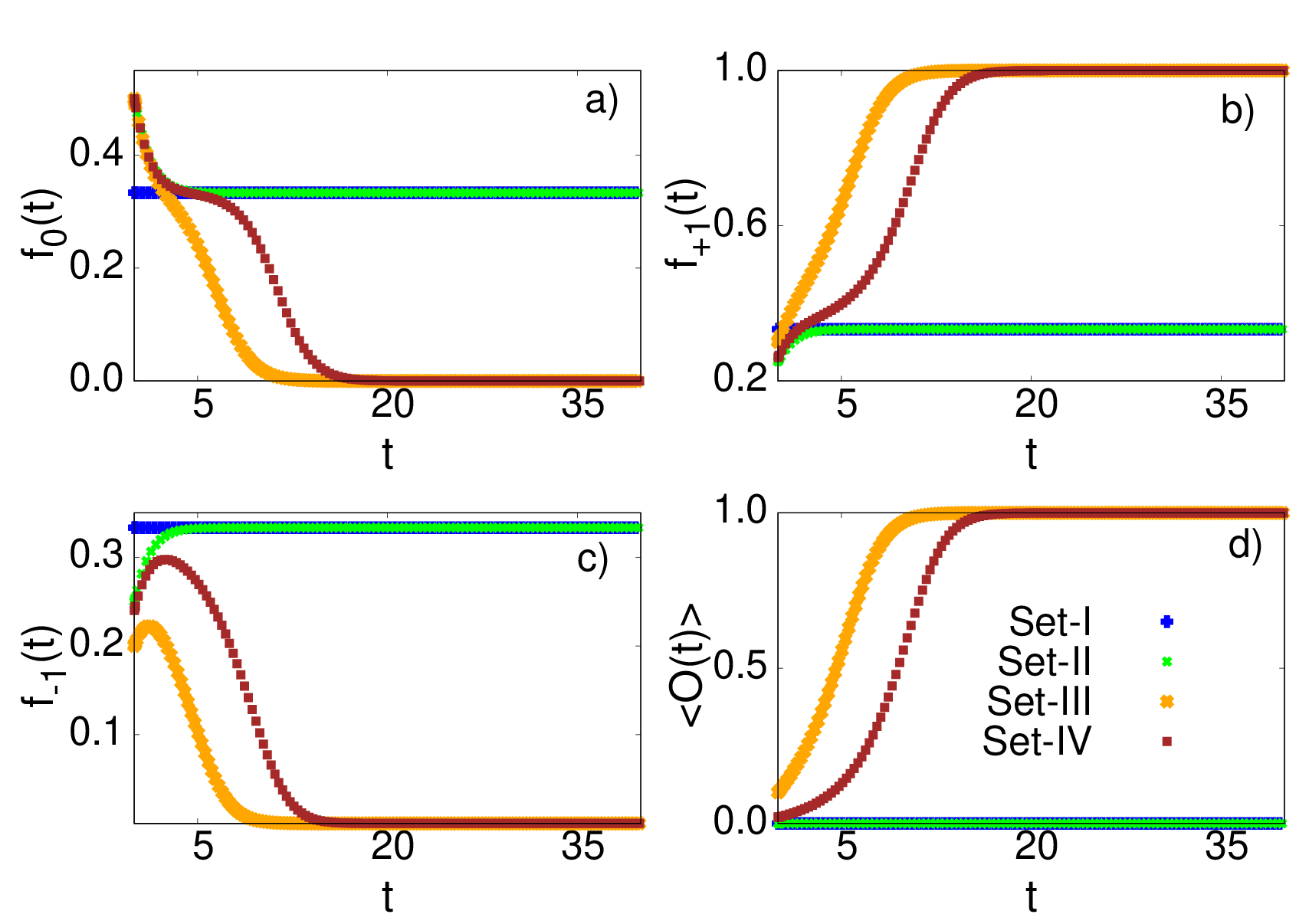}
\caption{Results for $q=0$ 
obtained from  analytical solution  using the initial configurations given  in Table \ref{table:1}.
The three densities  and the ensemble averaged order parameter are  
shown  as functions of time in  (a), (b), (c) and (d) respectively. The time evolution of sets I and II merge within a few steps as expected.} 
\label{fig:q=0,0-case-1-set-pic} 
\end{figure}

\begin{figure}[h]
\centering
\includegraphics[width=8cm,angle=0]{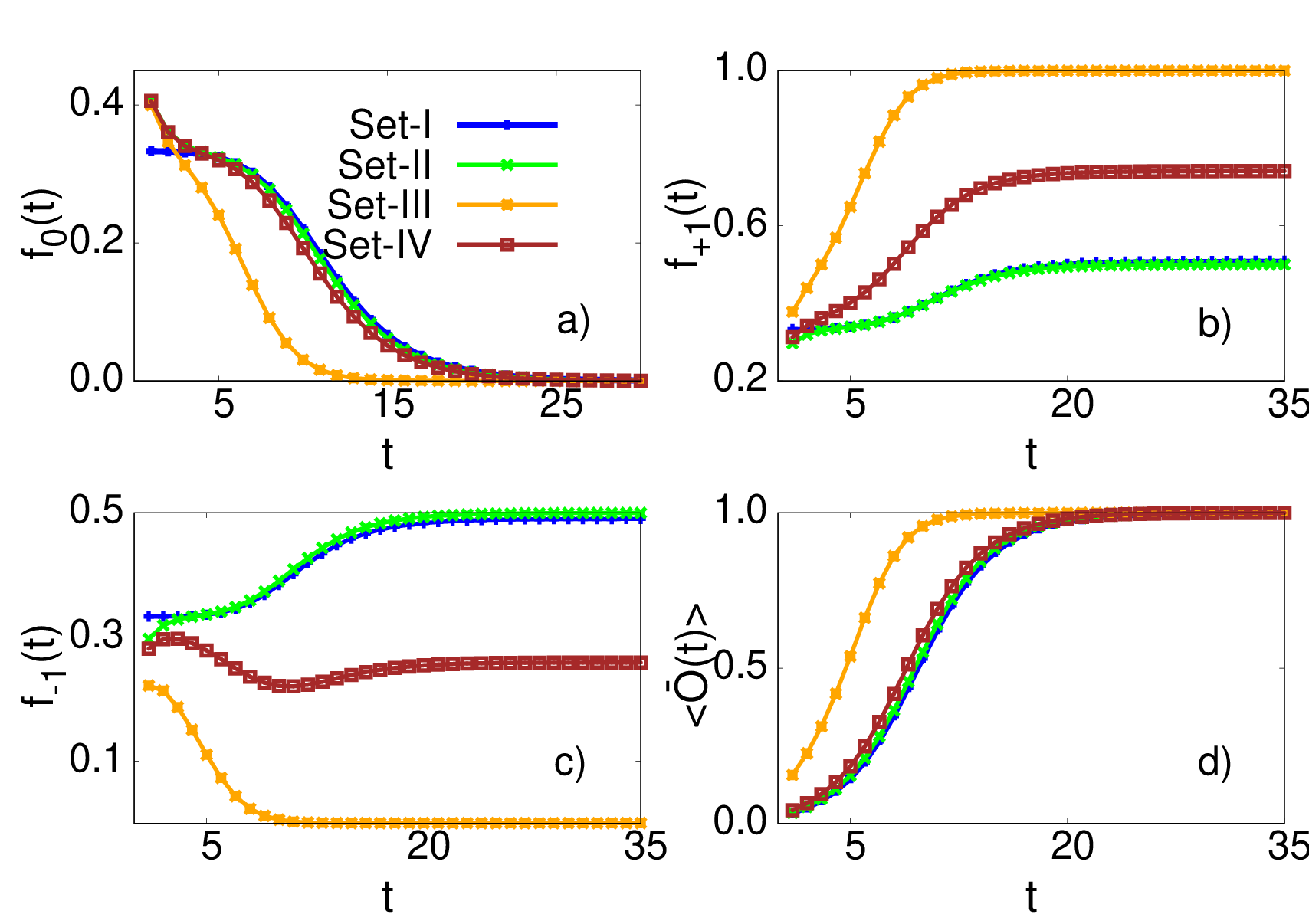}
\caption{Results for $q=0$ 
obtained from  Monte Carlo simulations ( of system size $N=2^{10}$ )  using the initial configurations given  in Table \ref{table:1}.
The three densities  and the ensemble averaged order parameter are  
shown  as functions of time in  (a), (b), (c) and (d) respectively.} 
\label{fig:q=0,0-case-1-set-simulation-pic} 
\end{figure}

\begin{figure}
\includegraphics[width=8cm]{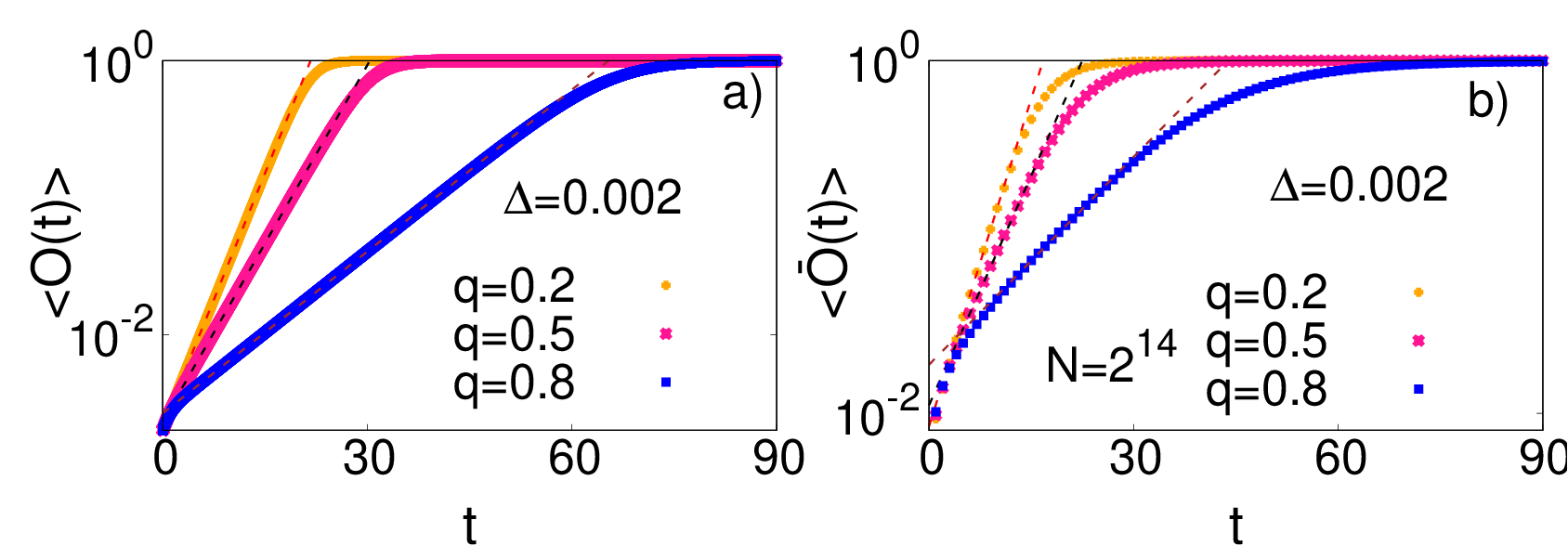}
\caption{ The order parameter versus time 
  variations for the initial condition $f_{\pm} = 1/3 \pm \Delta/2$ shown for different values of $q$ using (a) analytical method and  (b) numerical 
simulation. 
The best fitting curves for the growth shows an exponential  form $\exp [\beta(q)t]$. 
}
\label{orderpara2}
\end{figure}

\begin{figure}
\includegraphics[width=8cm]{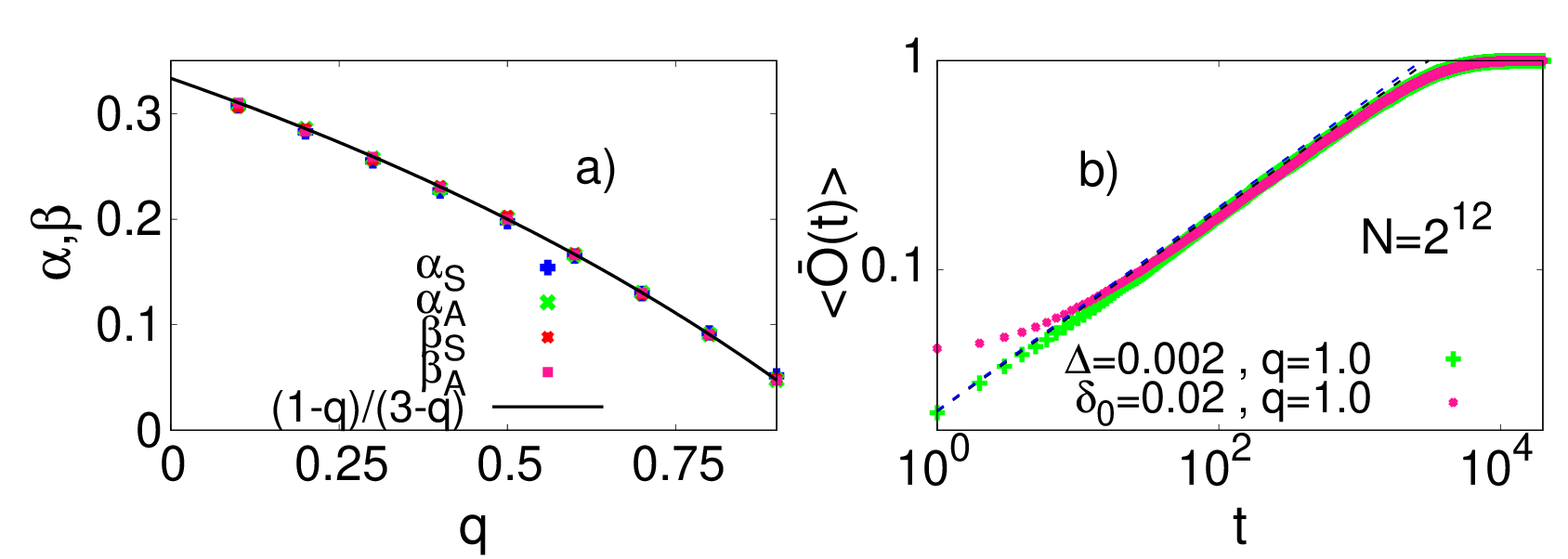}
\caption{(a) The values of $\alpha$ and $\beta$ (suffix A  and S denoting analytical and simulation results respectively) shown   against $q$   agree very well with the  analytical form given by eq. \ref{alpha}. (b) The order parameter against time for $q=1$ obtained using the numerical simulations is  shown to follow a power law 
behaviour with the exponent $= 0.49 \pm 0.01$. The initial conditions are the same as in Fig. \ref{orderpara1}b and \ref{orderpara2}b.}
\label{alphabeta}
\end{figure}

Let us next discuss the case for $q=0$, the other extreme limit.
This is the case when extreme switches are not allowed and is identical 
to the model considered in \cite{BCS} when  all interactions are positive and equal to one. In that case,  
an ordered state is expected at long times. However,  as already discussed in the last subsection,
the 
 $\Delta = 0$ point is the FFP here leading to 
 $\langle O(t) \rangle = 0$ for all $t$ when the time evolution is studied using the analytical equations.
The results are presented in Fig. \ref{fig:q=0,0-case-1-set-pic}.
We note that the  set I  does not evolve at all and for set II, the densities evolve 
before terminating at the FFP, 
 consistent with the analysis presented in the previous subsection.   
%
Initial states with non-zero  order  show that the system reaches a consensus state.  We also observe 
that if the initial state is close to a disordered state (set IV),  the system spends a longer time to reach consensus.   
Once again, in the numerical simulations  $\langle \bar O (t \to \infty) \rangle  = 1 $  for all initial configurations
as shown in Fig. \ref{fig:q=0,0-case-1-set-simulation-pic}. In the completely disordered  case, again random fluctuations  are responsible for driving an individual  system to a consensus state.

For other values of $q\neq 1$, the  qualitative behaviour of the time evolution is  similar to that of $q=0$. One gets consensus states starting from initially biased states.  The disordered states flow to the FFP when the time evolution is studied   
by solving the differential equations numerically,
as expected.  
Consensus is reached in individual configurations starting from  any initial state
 in the numerical simulations. 

We have already discussed the growth of the order parameter
for initially ordered states with small deviations from the FFP taken in a particular manner. 
 In this section we discussed  the time evolution using various 
other initial configuration. It is found that any  initially ordered state finally attains consensus and the 
the growth of the  order 
parameter is found to be exponential in all cases given by  
  $\exp[\beta (q) t]$. In particular we show in Fig. \ref{orderpara2} the case when the initial condition is  $f_{\pm} = 1/3 \pm  \Delta/2$ with $\Delta =  0.002$.   

The values of $\alpha(q)$  and $\beta(q)$ obtained from the numerical solution of the rate equations as well as using Monte Carlo simulations 
 are very close to each other  
as  shown in 
Fig. \ref{alphabeta}.   
So we  conclude that when the system orders
for  $q \neq 1$, 
the initial growth of the order parameter   is given by a unique 
exponential form,
 independent of the initial condition,   with the exponent given by 
eq. \ref{alpha}.

Since the magnitude of the order parameter increases, a steady state must  imply $f_0 =0$. 
With $f_0 = 0$, we have from equations \ref{f+equation} and \ref{f-equation} that in the steady state,  either  $f_{+1}$ or  $f_{-1}$ must be  zero (or unity). 
Hence  
the   consensus  state will be reached for all $q \neq 1$.

For $q=1$, although the analytical results show that the consensus states are  not reached in the  thermodynamic limit, 
for finite systems, we  find a unique behaviour for the growth of the order parameter from the numerical simulation. Instead of exponential, it  displays a slower power law variation with the exponent very close to 0.5.
The data are presented in Fig. \ref{alphabeta}b.

\subsection{Consensus times}

\begin{figure}[h]
\centering
\includegraphics[width=8cm]{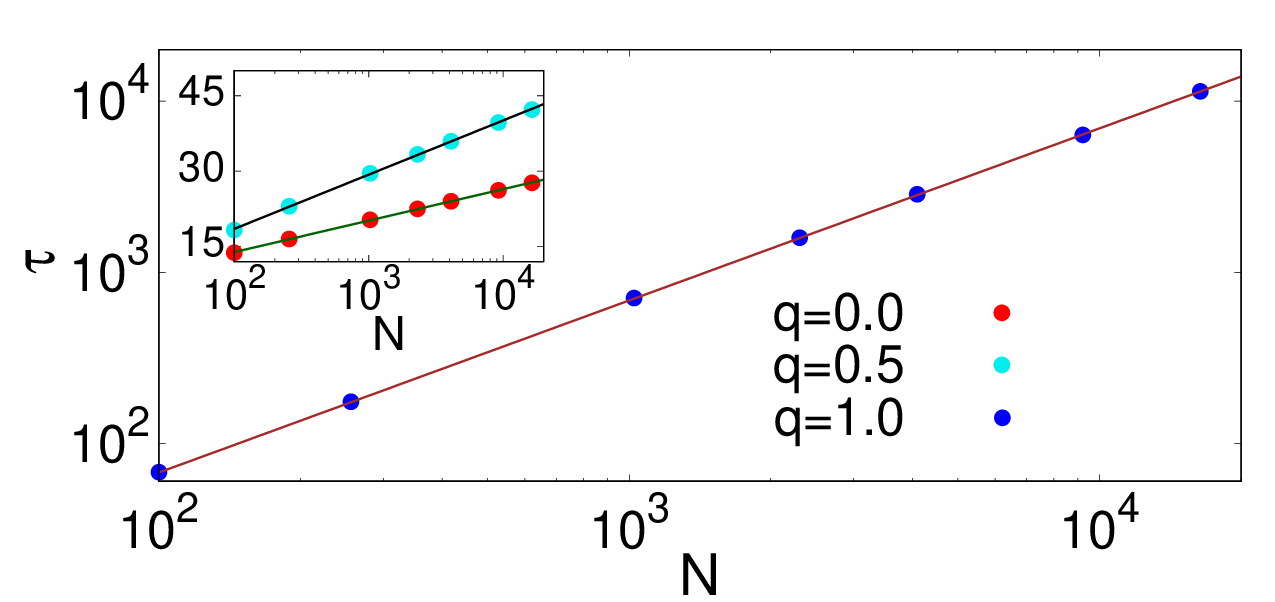}
\caption{Variation of average consensus time for different values of $q$. 
The main plot shows the $q=1$ results obtained using  system sizes up to  $N = 2^{14}$. Inset shows results for 
two values of $q\neq 1$ obtained in  system sizes $ \leq  2^{16}$ agents. } 
\label{fig:consensus-pic} 
\end{figure}

From the simulations, $\tau$, the average time to reach the consensus state has been estimated
for different system sizes.
Once again, we find different behaviour for $q\neq 1$ and $q=1$. 
For $q=1$, $\tau \propto N$, 
while for other values of $q$,  the consensus time $\tau$   depends logarithmically  on the system size, with  $\tau$ increasing with  $q$.  
Fig. \ref{fig:consensus-pic} shows the data.  

\subsection{Biased initial conditions and exit probability}

%
%
%

\begin{figure}[h]
\centering
\includegraphics[width=0.5\textwidth,angle=0]{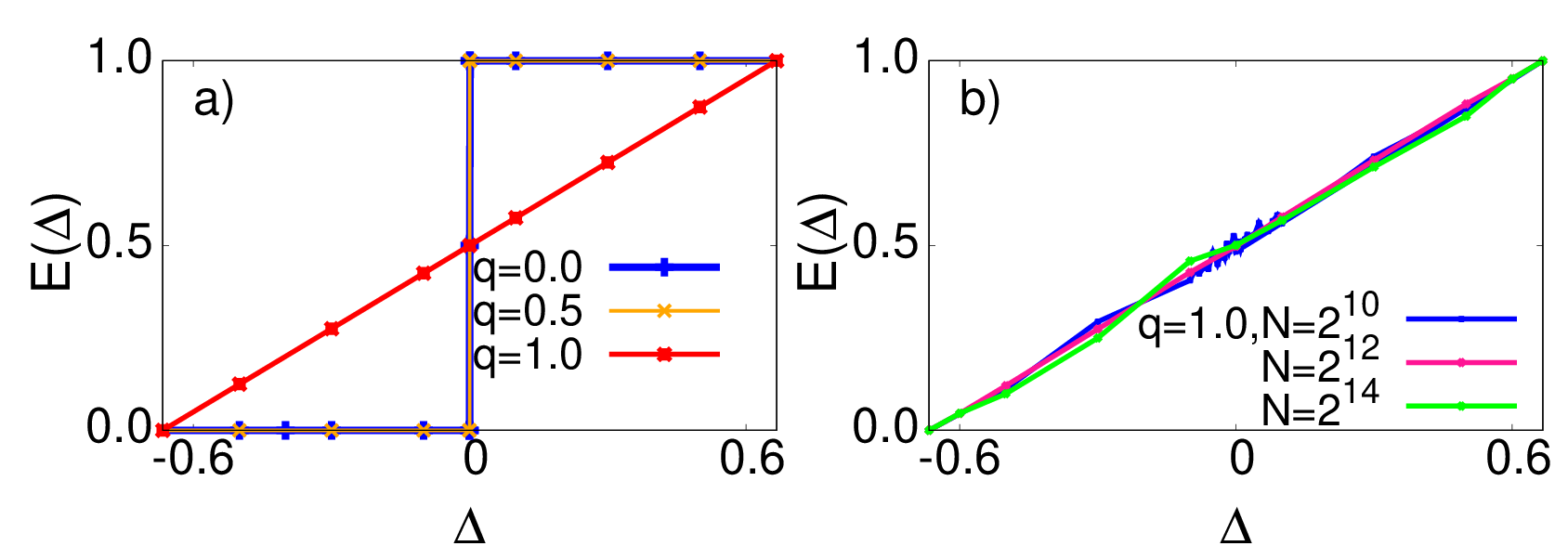}
\caption{ (a) $E(\Delta)$ against $\Delta$  obtained in the analytical 
method 
using eq. \ref{satexit}  
for different values of $q$  
(b) $E(\Delta)$ for $q=1$ using the numerical simulations for different system sizes shows a system-size independent linear behaviour.}
\label{fig:Osat-q-MF-pic} 
\end{figure}

\begin{figure}[h]
\centering
\includegraphics[width=0.5\textwidth,angle=0]{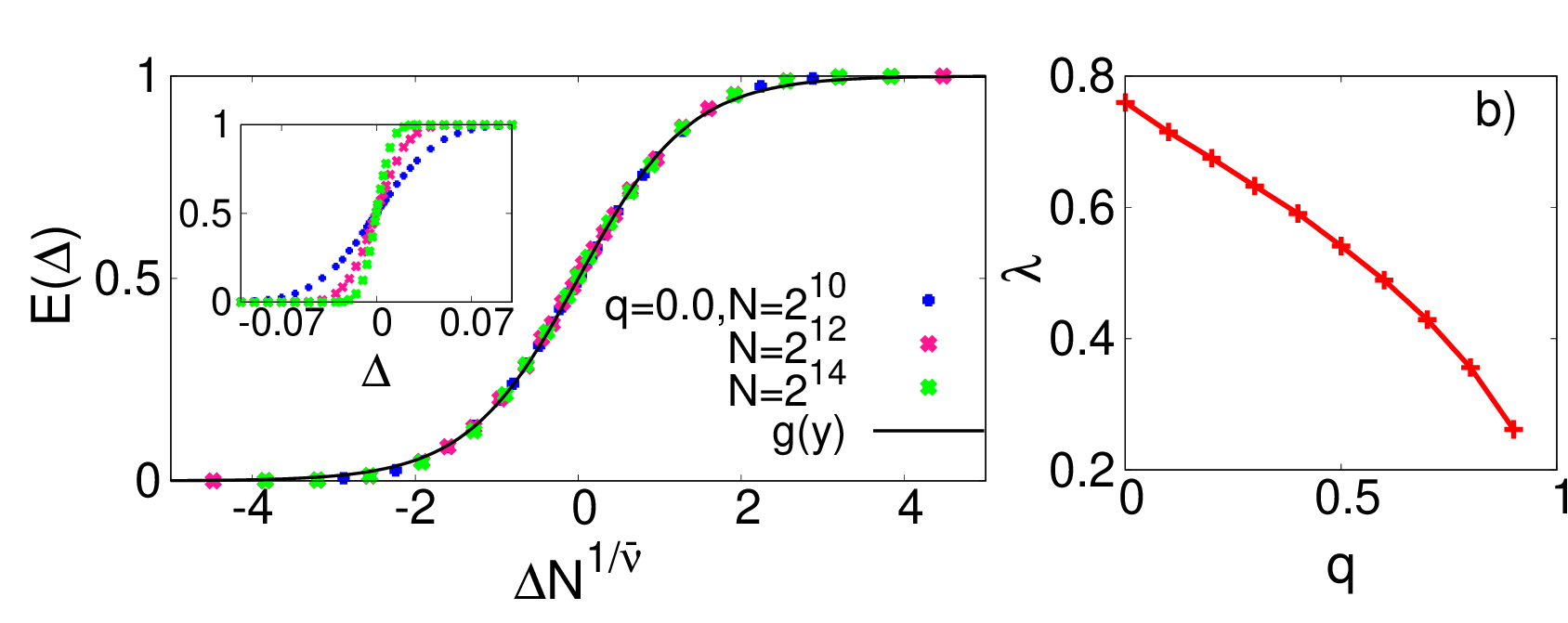}
\caption{(a) Data collapse  of $E(\Delta)$ for different system sizes   
is shown  against scaled $\Delta$ values for 
 $q=0$.  Inset  shows the raw data. 
(b) The variation of the parameter $\lambda$ against $q$.}
\label{fig:exit-probability-Q-MF-pic} 
\end{figure}

To calculate the exit probability, we consider 
 the   biased initial condition  $f_0 = 1/3$ and $f_{\pm 1} = 1/3 \pm  \Delta/2$. As already mentioned, we calculate the exit probability as a function of $\Delta$. 

The exit probability for $q=1$  has a completely different behaviour compared to other values of $q$. 
 We find that it has linear variation given by  $E(\Delta) = 1/2 +3\Delta/4$ shown in Fig. \ref{fig:Osat-q-MF-pic}a. 
A linear behaviour is expected in conserved systems but it is intriguing 
that even here, there is a linear behaviour although the system is  not exactly conserved. 
In this case, the simulations also agree 
as we take into account whether the consensus reached is for all $+1$ or all $-1$ states.    
Fig. \ref{fig:Osat-q-MF-pic}b shows the results are consistent with a linear variation of  $E(\Delta)$  when $q = 1$ and  shows that it 
is also  independent of the system size.

Analytical results for the exit probability, for $q \neq 1$, shows  a step function like behaviour;
 $E(\Delta) = 1$ for $\Delta > 0$ and equal to zero for $\Delta < 0$. The fixed point analysis also indicates 
$E(\Delta=0) = 1/2$. 
 Hence   any  
biased state with a majority  opinion equal to 1 (-1)  will end up with all opinions equal to 1 (-1).

The simulation results for $E(\Delta)$ for $q \neq 1$ shows strong finite size dependence. 
When plotted against $N^{1/\bar\nu} \Delta$, the data collapse in a single curve indicating   
\begin{equation}
E(\Delta)=g(N^{1/\bar \nu}\Delta),
\label{scale1}
\end{equation}
where $g$ is a scaling function. 
This is true for any value of $q \neq 1$ with a universal value of   $\bar \nu  \approx  2$.
Fig. \ref{fig:exit-probability-Q-MF-pic}a shows the data for $q=0$. 

The scaling function $g$ in eq. \ref{scale1} can be  approximated by 
\begin{equation}
g(y)=[1+\tanh(\lambda y)]/2, 
\end{equation}
as obtained earlier in a few other models  \cite{pr_sb_ps,pr_ps17,pm_ps17,sm_sb_ps,RRPS}. 
We find that   $\lambda$ decreases  as the value of $q$  increases from zero (Fig. \ref{fig:exit-probability-Q-MF-pic}b.).
 
 In the thermodynamic limit,  the exit probability shows the step function behaviour. However, 
for finite systems, it is $S$ shaped and from the finite size analysis we can conclude that  the range  of $\Delta$ over which it is neither zero or unity to a large extent, is inversely proportional to $\lambda N^{1/\bar \nu}$.

We end this section commenting that  the two different behaviour of the exit probability 
shown in Fig. \ref{fig:Osat-q-MF-pic}a  are analogous to the Ising Glauber model in dimensions greater than unity and voter model (in any dimension) respectively for $q \neq 1$ and $q=1$.

\section{Summary and Discussions}

In this paper, the 
evolution of the opinions in a kinetic exchange model has been studied 
using both analytical and numerical methods. The three discrete opinion values 
used here are quantised by $0, \pm 1$. 
The mean field  differential equations for the  rate of change of the population densities 
having the three opinions have been  derived and  analysed. Here, the parameter $q$ determines the value of the interaction $\mu$, a random variable, which can have binary values 1 and 2. When $\mu =2$, which occurs with a probability  $q$, there is 
a possibility that the opinion value switches from one extreme value to the other. The $q=0$ case, where $\mu$ can have a single value equal to unity,   has been considered earlier
in several studies in different contexts. 

Let us first summarise the main results obtained:\\
(a) Any initially ordered state will reach a consensus state for $q \neq 1$.\\
(b)  
 A frozen disordered fixed point exists; all initially disordered states flow there.\\
(c) The growth of the order parameter is exponential for $q \neq 1$.\\ 
(d) A quasi-conservation exists for $q \neq 1$ leading to different saturation behaviour of the order parameter and exit probability.

The results are qualitatively different for $q=1$ and $q\neq 1$. The analytical solution, which is valid in the thermodynamic limit,
shows that 
 for $q=1$ 
the dynamics are  quasi-conservative as the order parameter remains constant after a very short transient time.
This indicates that the system does not order fully for any initial configuration with initial
order parameter less than 1.  The linear behaviour of the exit probability is similar to 
what is seen for a conservative  dynamics as for example in  the Voter model in all dimensions and the Ising Glauber model in one dimension. This is actually quite interesting, as the present model
does not  strictly conserve the order parameter; the saturation value is 
not exactly equal to the    initial one.  
But the linear behaviour of the exit probability can still 
occur  if  the saturation value of the order parameter varies linearly with the initial value which we have checked to be true here. 

The  $q=1$ case is in fact very similar to the Voter model;  as $f_0$  goes to zero very fast, 
it effectively renders the  system to a 
 binary opinion model within a short time scale  with the transition rates identical  to those in the Voter model \cite{bennaim}. Like the voter model, here the agent 
 adapts the opinion of the other agent with whom she interacts irrespective 
of her own opinion. 
We also obtain the result that the average consensus time is proportional to $N$ for $q=1$, a result valid for the mean field voter model. 

In the analytical approach, one  essentially obtains the ensemble averages  in  the thermodynamic limit. 
 Initial configurations with nonzero order will eventually reach the consensus state 
for $q \neq 1$. 
We also find that  this growth behaviour is unique, i.e., does not depend on the initial state but only on $q$. This is not surprising as
it is expected that there will be a single time scale in the system. Such exponential growths have been recently 
observed in the mean field Ising model with finite coordination number also \cite{RRPS}. 

The analytical approach also leads to the interesting result that initially disordered state, that can be realised in many ways,  will flow towards the so called frozen fixed point 
at a   rate  independent of $q$.  
In comparison, in binary models like the Ising model, the disordered state is unique, 
characterised by exactly  half of the  relevant degrees of freedom  
belonging to one state. Hence no such flow can be observed there.

In the numerical simulations, one can keep track of the individual configurations.
For all $q$  values we get a consensus state finally  for states  starting from 
partially disordered states.  For $q\neq 1$, this is the same result obtained from analytical 
treatment. However it is not expected that consensus will be 
obtained   for $q=1$  for any initial state  and 
for $q\neq 1$,   
for initially fully disordered configurations.  
This contradictory result obtained in the simulations is argued to be due to  finite size effects. 
In finite systems,  random fluctuations can drive
the system to a consensus state (which implies that the absolute value of the order parameter is unity) even if the initial configuration is 
  fully disordered. 
This has been observed in spin models also, e.g., in the one dimensional Ising Glauber model
for which the ensemble averaged order parameter is conserved but still consensus states 
can be reached in numerical simulations starting from disordered states. 
Numerical simulations also show that for $q=1$, the growth follows a  power law  behaviour, which is much slower than exponential. 
As a result,  the consensus time 
is linear in $N$ for $q=1$,  
compared to the weak logarithmic dependence on the system size when $q \neq 1$.

The  exit probability for $q \neq 1$ indicates a step function behaviour in the
thermodynamic limit. It shows strong finite size effects as indicated from the numerical simulations. 
As observed in some other models, a scaling behaviour is obtained dictated by  two parameters $\bar \nu$ and $\lambda$.
The value of $\bar \nu$ is independent of $q$, a result similar to that  in several other models where also $\bar \nu$ does not depend on the
model parameter. 
However, the value of $\bar \nu \approx 2$ is clearly different from the ones found earlier for Ising-like and other opinion dynamics models
\cite{pr_sb_ps,pr_ps17,pm_ps17,sm_sb_ps,RRPS}. 
$\lambda$, on the other hand is dependent on the parameter $q$, which was also found to be true in the
other models. 
The linear variation of the exit probability in the $q=1$ case, independent of system size, also indicates that one will get minority spreading here \cite{galam}.


In conclusion, the present results indicate that 
 a society  attains stability when  people have less influence on others, i.e., $q$ is small, with the consensus state attained very fast. Essentially, the $q \neq 1$ model is qualitatively similar to the $q=0$ model, with a $q$ dependent timescale to reach the consensus state which diverges as $1/\alpha \propto  (1-q)^{-1}$. So the extreme switches cause a delay in reaching the consensus 
as they increase in number.  
 $q=1$, which allows the maximum possible switches between extreme opinion values, essentially leads to a fragmented society. That this does not happen  usually  signifies that real systems may be mimicked by a
 $q \neq 1$ value in this model. It also shows that an initially disordered society will remain so when one 
considers the ensemble average, however, individual configurations do reach consensus. From the perspective of statistical physics, we have presented a model with a rich behaviour as $q$ is changed;  at $q=1$ a Voter model-like behaviour is seen that changes to  a finite dimensional 
spin-model like behaviour for any $q < 1$.  
As future studies, it will be interesting to consider negative interactions between the agents which  will introduce a noise that can drive a  order disorder transitions. This will also make it 
closer to reality and would open up the possibility to compare with time dependent real data. 
 Another interesting possibility is to consider general opinion values instead of $\pm 1, 0$ \cite{hadz,gekle,galam3} and introduce transition between 
any two states and see how it compares with the present case.

%
%

\medskip

{\bf {Acknowledgement}} 

The authors acknowledge very useful discussions with Soumyajyoti Biswas. 
PS acknowledges financial support from SERB (Government of India) through scheme no MTR/2020/000356.

\end{document}